\input{aipcheck}

\documentclass[
    ,final            
    ,numberedheadings 
  ]{aipproc}

\layoutstyle{6x9}

\usepackage{amsfonts}
\usepackage{amssymb}
\usepackage{graphicx} 
\usepackage{color}

\definecolor{darkgreen}{rgb}{0,0.5,0}
\definecolor{purple}{rgb}{0.5,0,0.5}
\definecolor{nblue}{rgb}{0.0,0.0,0.50}
\definecolor{scarlet}{rgb}{1.0,0.2,0}
\usepackage[colorlinks=true, pdfstartview=FitV, linkcolor=purple, citecolor= purple, urlcolor=blue]{hyperref}

\newcommand{\sfrac}[2]{\mbox{\footnotesize $\displaystyle \frac{#1}{#2}$}} 
\newcommand{\lsim}{\mathrel{\rlap{\lower4pt\hbox{\hskip0pt$\sim$}} 
\raise1pt\hbox{$<$}}}           
\newcommand{\gsim}{\mathrel{\rlap{\lower4pt\hbox{\hskip0pt$\sim$}} 
\raise1pt\hbox{$>$}}}           

\begin{document}

\title{A Perspective on Hadron Physics}



\author{Arne H\"oll}{
  address={Institut f\"ur Physik, Universit\"at Rostock, D-18051 Rostock, Germany}
}

\author{Craig D.\ Roberts}
{
address={Institut f\"ur Physik, Universit\"at Rostock, D-18051 Rostock, Germany},
altaddress={Physics Division, Argonne National Laboratory, Argonne IL 60439, 
USA}
}

\author{Stewart V.\ Wright}{
  address={Physics Division, Argonne National Laboratory, Argonne IL 60439, 
USA}
}

\begin{abstract}
The phenomena of confinement and dynamical chiral symmetry breaking are basic to understanding hadron observables.  They can be explored using Dyson-Schwinger equations.
The existence of a systematic, nonperturbative and symmetry preserving truncation of these equations enables the proof of exact results in QCD, and their illustration using simple but accurate models.  We provide a sketch of the material qualitative and quantitative success that has been achieved in the study of pseudoscalar and vector mesons.  Efforts are now turning to the study of baryons, which we exemplify via a calculation of nucleon weak and pionic form factors.
\end{abstract}

\maketitle


\section{Introduction}
We begin with a key question: how does one make an almost massless bound state from two massive constituents?  Naturally, the bound state is the pion and the massive components are constituent-quarks.  It has long been known \cite{gmor68} that 
\begin{equation}
\label{gmorS}
m_\pi^2 \propto m_q\,,
\end{equation}
where $m_q$ is the light-quark current-mass that appears in QCD's Lagrangian.  While it is possible to construct a quantum mechanical model with a potential finely tuned to give a massless pseudoscalar bound state composed of heavy constituents, in such a framework $m_\pi \propto M_q$, where $M_q$ is the constituents' mass.  This is plainly not the way to a veracious understanding of strong interaction physics.  

True comprehension of the visible universe requires that we learn just what it is about QCD which enables the formation of an unexpectedly light pseudoscalar meson from two rather massive constituents.  The correct understanding of hadron observables must explain why the pion is light but the $\rho$-meson and the nucleon are heavy.  The keys to this puzzle are QCD's \textbf{emergent phenomena}: \textit{confinement} and \textit{dynamical chiral symmetry breaking} (DCSB).  Confinement is the feature that no matter how hard one strikes a hadron, it never breaks apart into quarks and/or gluons that reach a detector.  DCSB is signalled by an apparently unnatural pattern of bound state masses in the strong interaction spectrum, and can only be fathomed once one grasps the nature of a well-defined and valid chiral limit.  Thereafter can follow an understanding of the connection between a current-quark and a constituent-quark, and subsequently Eq.\,(\ref{gmorS}).  QCD's emergent phenomena are not apparent in the action.  Yet they are the dominant determining characteristics of hadron properties.  Attaining an understanding of these phenomena is one of the greatest intellectual challenges in physics.

A nonperturbative method for solving quantum field theory is necessary in order to answer the question we have posed, and those which shall follow.  The Dyson-Schwinger equations (DSEs) are one such tool.  At the simplest level the DSEs provide a generating tool for perturbation theory and, because QCD is asymptotically free, this means that any model-dependence in their application can be restricted to the infrared (long-range) domain.  The solutions of the DSEs are Schwinger functions (Euclidean space Green functions) and because all cross-sections can be constructed from such $n$-point functions the DSEs can be used to make predictions for real-world experiments.  In this mode they provide a means by which to use nonperturbative strong interaction phenomena to map out, e.g., the behaviour at long-range of the interaction between light-quarks.  A nonperturbative solution of the DSEs enables the study of: hadrons as composites of dressed-quarks and -gluons; the phenomena of confinement and DCSB; and therefrom an articulation of any connection between them.  One of the merits of this is that any assumptions employed, or guesses made, can be tested, verified and improved, or rejected in favour of more promising alternatives.  The modern application of these methods is described in Refs.\,\cite{bastirev,cdrwien,reinhardrev,pieterrev}, while Ref.\,\cite{HUGS} provides a pedagogical overview.

The DSEs are a countable infinity of equations, which are vitally important in proving the renormalisability of quantum field theories.  However, the coupling between equations is at the heart of a persistent challenge to their application.  This relationship means that in order to arrive at a tractable problem one must employ a truncation.  Perturbation theory is ever popular.  However, it is not useful in connection with the nonperturbative phenomena that provide the keystones of hadron physics.  Fortunately, at least one systematic, nonperturbative and symmetry preserving truncation of the DSEs exists \cite{munczek,truncscheme}.  This enables the proof of exact results using the DSEs.  Moreover, that the truncation scheme is also tractable provides a method by which the exact results may be illustrated and, furthermore, a practical tool for the prediction of observables that are accessible at contemporary experimental facilities.  The consequent opportunities for rapid feedback between experiment and theory brings within reach an intuitive understanding of nonperturbative strong interaction phenomena.

\section{Gap Equation}
The renormalised gap equation in QCD may be written
\begin{eqnarray}
S(p)^{-1} & =&  Z_2 \,(i\gamma\cdot p + m^{\rm bm}) + \Sigma(p)\,, \label{gendse} \\
\Sigma(p) & = & Z_1 \int^\Lambda_q\! g^2 D_{\mu\nu}(p-q) \frac{\lambda^a}{2}\gamma_\mu S(q) \Gamma^a_\nu(q,p) , \label{gensigma}
\end{eqnarray}
where $\int^\Lambda_q$ represents a Poincar\'e invariant regularisation of the integral, with $\Lambda$ the regularisation mass-scale \cite{mrt98,mr97}, $D_{\mu\nu}(k)$ is the dressed-gluon propagator, $\Gamma_\nu(q,p)$ is the dressed-quark-gluon vertex, and $m^{\rm bm}$ is the $\Lambda$-dependent current-quark bare mass.  The quark-gluon-vertex and quark wave function renormalisation constants, $Z_{1,2}(\zeta^2,\Lambda^2)$, depend on the renormalisation point, $\zeta$, the regularisation mass-scale and the gauge parameter.  The gap equation's solution has the form 
\begin{eqnarray} 
 S(p) & = & \frac{1}{i \gamma\cdot p \, A(p^2,\zeta^2) + B(p^2,\zeta^2)} = \frac{Z(p^2,\zeta^2)}{i\gamma\cdot p + M(p^2)} \,.
\label{Sgeneral}
\end{eqnarray} 
It is obtained from Eq.\,(\ref{gendse}) augmented by the renormalisation condition
\begin{equation}
\label{renormS} \left.S(p)^{-1}\right|_{p^2=\zeta^2} = i\gamma\cdot p +
m(\zeta)\,,
\end{equation}
where $m(\zeta)$ is the renormalised mass: 
\begin{equation}
Z_2(\zeta^2,\Lambda^2) \, m^{\rm bm}(\Lambda) = Z_4(\zeta^2,\Lambda^2) \, m(\zeta)\,,
\end{equation}
with $Z_4$ the Lagrangian mass renormalisation constant.  In QCD the chiral limit is strictly and unambiguously defined by
\begin{equation}
\label{limchiral}
Z_2(\zeta^2,\Lambda^2) \, m^{\rm bm}(\Lambda) \equiv 0 \,, \forall \Lambda \gg \zeta \,,
\end{equation}
which states that the renormalisation-point-invariant current-quark mass $\hat m = 0$.

In the absence of interactions $Z(p^2)=1$ and $M(p^2) = m_q$ in Eq.\,(\ref{Sgeneral}).  On the other hand, the behaviour of these functions in QCD is a longstanding prediction of DSE studies \cite{cdragw}, which could have been anticipated from Refs.\,\cite{lane,politzer}: the functions receive strong momentum-dependent corrections at infrared momenta so that $Z(p^2)$ is suppressed and $M(p^2)$ enhanced.  These DSE predictions are confirmed in numerical simulations of lattice-QCD \cite{bowman2}, and the conditions have been explored under which pointwise agreement between DSE results and lattice simulations may be obtained \cite{bhagwat,alkoferdetmold}.

The gap equation's kernel, Eq.\,(\ref{gensigma}), is constructed from the contraction of the dressed-gluon two-point function and the dressed-quark-gluon vertex.  In Landau gauge 
\begin{equation}
\label{gluonZ}
D_{\mu\nu}(p)= \left(\delta_{\mu\nu} - \frac{p_\mu p_\nu}{p^2}\right) \frac{F(p^2,\zeta^2)}{p^2}\,.
\end{equation}
The modern DSE perspective on $F(p^2,\zeta^2)$ is reviewed in Ref.\,\cite{reinhardrev}: these studies predicted that $F(p^2)$ is suppressed at small $p^2$; i.e., in the infrared, with the deviation from expectations based on perturbation theory becoming apparent at $p^2 \simeq 1\,$GeV$^2$.  A mass-scale of this magnitude has long been anticipated as characteristic of nonperturbative gauge-sector dynamics and its origin is fundamentally the same as that of $\Lambda_{\rm QCD}$, which appears in perturbation theory.  These DSE predictions, too, have been verified in contemporary simulations of lattice-regularised QCD \cite{latticegluon}.

The remaining piece of the gap equation's kernel is the dressed-quark-gluon vertex, whose form is the subject of contemporary research.  In correlating lattice-QCD results on the dressed-quark and -gluon propagators via the gap equation it was found \cite{bhagwat} that the vertex must exhibit an infrared enhancement.  This was anticipated in Ref.\,\cite{hawes} and confirmed in Ref.\,\cite{alkoferdetmold}.  The exact nature of this enhancement and its origin in QCD is currently being explored; e.g., Refs.\,\cite{jisvertex,bhagwatvertex,bhagwat2}.

\section{Mesons}
Dyson-Schwinger equation studies have established a reliable picture of key propagators and vertices in QCD.  It is now natural to ask: what about bound states?  Without them, of course, a direct comparison with experiment is impossible.  Bound states appear as pole contributions to colour-singlet Schwinger functions and this observation may be viewed as the origin of the Bethe-Salpeter equation (BSE).  

The DSE for the dressed-quark-gluon vertex can be viewed as a BSE.  So can that for the dressed-quark-photon vertex.  The latter is a colour singlet vertex and its lowest mass pole-contribution is the $\rho$-meson \cite{marisphotonvertex}.  This fact underlies the success of $\rho$-meson dominance phenomenology.  

The axial-vector vertex is of primary interest to hadron physics.  It may be obtained as the solution of the inhomogeneous Bethe-Salpeter equation
\begin{equation}
\label{avbse}
\left[\Gamma_{5\mu}(k;P)\right]_{tu}
 =  Z_2 \left[\gamma_5\gamma_\mu\right]_{tu} + \int^\Lambda_q
[\chi_{5\mu}(q;P)]_{sr} K_{tu}^{rs}(q,k;P)\,,
\end{equation}
where $\chi_{5\mu}(q;P)=S(q_+) \Gamma_{5\mu}(q;P) S(q_-)$, $q_\pm = q \pm P/2$, and the colour-, Dirac- and flavour-matrix structure of the elements in the equation is denoted by the indices $r,s,t,u$.  In Eq.\,(\ref{avbse}), $K(q,k;P)$ is the fully-amputated quark-antiquark scattering kernel.  It is one-particle-irreducible and hence does not contain quark-antiquark to single gauge-boson annihilation diagrams, such as would describe the leptonic decay of the pion, nor diagrams that become disconnected by cutting one quark and one antiquark line.  If one knows the form of $K$ then one \emph{completely} understands the nature of the interaction between quarks in QCD.

\medskip

%
\hspace*{-\parindent}\underline{\textit{Model-independent results}}\\[1ex]
In quantum field theory, chiral symmetry and the pattern by which it is broken are expressed via the chiral Ward-Takahashi identity ($k_\pm = k \pm P/2$):
\begin{equation}
\label{avwtim}
P_\mu \Gamma_{5\mu}^H(k;P)  = \check{S}(k_+)^{-1} i \gamma_5\frac{T^H}{2}
+  i \gamma_5\frac{T^H}{2} \check{S}(k_-)^{-1}
- i\,\{ {M}^\zeta ,\Gamma_5^H(k;P) \} ,
\end{equation}
where the pseudoscalar vertex satisfies
\begin{equation}
\label{genpve}
\left[\Gamma_{5}^H(k;P)\right]_{tu} =
Z_4\,\left[\gamma_5 \frac{T^H}{2}\right]_{tu} \,+
\int^\Lambda_q \,
\left[ \chi_5^H(q;P)\right]_{sr}
K^{rs}_{tu}(q,k;P)\,,
\end{equation}
with
$\check{S}= {\rm diag}[S_u,S_d,S_s,\ldots]$ 
and $M ^\zeta = {\rm diag}[m_u(\zeta),m_d(\zeta),m_s(\zeta),\ldots]$.  We have written Eqs.\,(\ref{avwtim}), (\ref{genpve}) for the case of a flavour-nonsinglet vertex in a theory with $N_f$ quark flavours.  The matrices $T^H$ are constructed from the generators of $SU(N_f)$ with, e.g., \mbox{$T^{\pi^+}=\mbox{\small $\frac{1}{2}$} (\lambda^1+i\lambda^2)$} providing for the flavour content of a positively charged pion.  

The axial-vector Ward-Takahashi identity relates the solution of a BSE to that of the gap equation.  If the identity is always to be satisfied and in a model-independent manner, as necessary in order to preserve an essential symmetry of the strong interaction and its breaking pattern, then the kernels of the gap and Bethe-Salpeter equations must be intimately related.  Any truncation or approximation of these equations must preserve that relation.  This is an extremely tight constraint.  Perturbation theory is one truncation that, order by order, guarantees Eq.\,(\ref{avwtim}).  However, perturbation theory is inadequate in the face of QCD's emergent phenomena.  Something else is needed. 

That need is satisfied by the systematic, nonperturbative and symmetry preserving truncation of the DSEs explained in Refs.\,\cite{munczek,truncscheme,bhagwatvertex,detmoldvertex}.  It enables a proof of Goldstone's theorem in QCD \cite{mrt98}.  Namely, in the chiral limit, Eq.\,(\ref{limchiral}), and with chiral symmetry dynamically broken: the axial-vector vertex, Eq.\,(\ref{avbse}), is dominated by the pion pole for $P^2\sim 0$ and the homogeneous, isovector, pseudoscalar BSE has a massless ($P^2 = 0$) solution.  The converse is also true, so that DCSB is a sufficient and necessary condition for the appearance of a massless pseudoscalar bound state of dynamically-massive constituents, which dominates the axial-vector vertex for infrared total momenta. 

Furthermore, from the axial-vector Ward-Takahashi identity and the existence of a systematic, nonperturbative symmetry-preserving truncation, one can prove the following identity involving the mass-squared of a pseudoscalar meson \cite{mrt98}:
\begin{equation}
\label{gengmor}
f_H \, m_H^2 = 
\rho_H(\zeta) {M}_H^\zeta,
\end{equation}
where ${M}_H^\zeta = m_{q_1}(\zeta) + m_{q_2}(\zeta)$ is the sum of the
current-quark masses of the meson's constituents;
\begin{equation}
\label{fH}
f_H \, P_\mu = Z_2 {\rm tr} \int_q^\Lambda \! \sfrac{1}{2} (T^H)^T \gamma_5
\gamma_\mu \check{S}(q_+)\, \Gamma^H(q;P)\, \check{S}(q_-)\,,
\end{equation}
where $(\cdot)^{\rm T}$ indicates matrix transpose, the trace is over all matrix indices; and
\begin{equation}
\label{qbqH}
\rho_H(\zeta) =   Z_4\, {\rm tr}\int_q^\Lambda \!
\sfrac{1}{2}(T^H)^T \gamma_5 \check{S}(q_+) \,\Gamma^H(q;P)\, \check{S}(q_-)=: \frac{-\langle \bar q q \rangle^H_\zeta}{f_H}  \,.
\end{equation}
The renormalisation constants in Eqs.\,(\ref{fH}), (\ref{qbqH}) play a pivotal role because the expressions would be meaningless without them.  They serve to guarantee that the quantities described are gauge invariant, and finite as the regularisation scale is removed to infinity.  Moreover, $Z_2$ in Eq.\,(\ref{fH}) and $Z_4$ in Eq.\,(\ref{qbqH}) ensure that both $f_H$ and the product $\rho_H(\zeta) M_H^\zeta$ are renormalisation point independent, which is an absolute necessity for any observable quantity.

Taking note that in a Poincar\'e invariant theory a pseudoscalar meson Bethe-Salpeter amplitude assumes the form
\begin{equation}
\Gamma_{H}^j(k;P) = T^H \gamma_5
\left[ i E_H(k;P)  + \gamma\cdot P \, F_H(k;P)+ \, \gamma\cdot k \,k\cdot P\, G_H(k;P)+
 \sigma_{\mu\nu}\,k_\mu P_\nu \,H_H(k;P) \right], \label{genpvv}
\end{equation}
then, in the chiral limit, one can also prove 
\begin{equation}
\begin{array}{ll}
f_H^0 E_H(k;0)  =   B(k^2)\,, & 
F_R(k;0) +  2 \, f_H^0 F_H(k;0)  = A(k^2)\,,\\
H_R(k;0) +  2 \,f_H^0 H_H(k;0)    = 0 \,, &
G_R(k;0) +  2 \,f_H^0 G_H(k;0)   =2 A^\prime(k^2)\,,
\end{array}
\label{bwti}
\end{equation}
where $f_H^0$ is the chiral limit value from Eq.\,(\ref{fH}), which is nonzero when chiral symmetry is dynamically broken.  The functions $F_R$, $G_R$, $H_R$ are associated with terms in the axial-vector vertex that are regular in the neighbourhood of $P^2 +m_H^2 = 0$ and do not vanish at $P_\mu = 0$.  These four identities are quark-level Goldberger-Treiman relations for the pion.  They are: exact in QCD; and a pointwise expression of Goldstone's theorem.  These identities relate the pseudoscalar meson Bethe-Salpeter amplitude directly to the dressed-quark propagator, Eq.\,(\ref{Sgeneral}).  The first explains why DCSB and the appearance of a Goldstone mode are so intimately connected, and the remaining three entail that in general a pseudoscalar meson Bethe-Salpeter amplitude has what might be called pseudovector components; namely: $F_H$, $G_H$, $H_H$.  It is the latter which, in a covariant treatment, guarantee that the electromagnetic pion form factor behaves as $1/Q^2$ at large spacelike momentum transfer \cite{mrpion}.

Equation (\ref{gengmor}) and its corollaries are fundamental in QCD.  To exemplify we'll focus first on the chiral limit behaviour of Eq.\,(\ref{qbqH})
whereat, using Eqs.\,(\ref{genpvv}) \& (\ref{bwti}), one finds
\begin{equation}
f_H^0\,\rho^0_H(\zeta) = 
Z_4(\zeta,\Lambda)\, N_c\, {\rm tr}_{\rm D} \int^\Lambda_q  S_{\hat m =0}(q) = - \langle \bar q q \rangle_\zeta^0  \,. \label{qbq0}
\end{equation}
Equation (\ref{qbq0}) is unique as the expression for the chiral limit \textit{vacuum quark condensate}.  It thus follows from Eqs.\,(\ref{gengmor}) \& (\ref{qbq0}) that in the neighbourhood of the chiral limit
\begin{equation}
\label{gmor}
(f_H^0)^2 \, m_H^2 = - \, M_H^\zeta\, \langle \bar q q \rangle_\zeta^0 +
{\rm O}(\hat M^2)\,.
\end{equation}
Hence Eq\,(\ref{gmorS}), which is commonly known as the Gell-Mann--Oakes--Renner relation, is a \textit{corollary} of Eq.\,(\ref{gengmor}).

Let's now consider another extreme; viz., when one of the constituents is a heavy quark, a domain on which Eq.\,(\ref{gengmor}) is equally valid.  In this instance Eq.\,(\ref{fH}) yields the model-independent result \cite{marismisha} 
\begin{equation}
\label{fHheavy}
f_H \propto \frac{1}{\sqrt{M_H}}\,;
\end{equation}
i.e., it reproduces a well-known consequence of heavy-quark symmetry \cite{neubert93}.  A similar analysis of Eq.\,(\ref{qbqH}) gives a new result \cite{marisAdelaide,mishaSVY}
\begin{equation}
\label{qbqHheavy}
- \langle \bar q q \rangle^H_\zeta = \mbox{constant} +
  O\left(\frac{1}{m_H}\right) \mbox{~for~} \frac{1}{m_H} \sim 0\,.
\end{equation}
Combining Eqs.\,(\ref{fHheavy}), (\ref{qbqHheavy}), one finds \cite{marisAdelaide,mishaSVY}
\begin{equation}
m_H \propto \hat m_f \;\; \mbox{for} \;\; \frac{1}{\hat m_f} \sim 0\,,
\end{equation}
where $ \hat m_f$ is the renormalisation-group-invariant current-quark mass of the pseudoscalar meson's heaviest constituent.  This is the result one would have anticipated from constituent-quark models but here we have indicated a direct proof in QCD. 

Pseudoscalar mesons hold a special place in QCD and there are three states, composed of $u$,$d$ quarks, in the hadron spectrum with masses below $2\,$GeV \cite{pdg}: $\pi(140)$; $\pi(1300)$; and $\pi(1800)$.  Of these, the pion [$\pi(140)$] is naturally well known and much studied.  In the context of a model constituent-quark Hamiltonian, these mesons are often viewed as the first three members of a $Q\bar Q$ $n\, ^1\!S_0$ trajectory, where $n$ is the principal quantum number; i.e., the $\pi(140)$ is viewed as the $S$-wave ground state and the others are its first two radial excitations.  By this reasoning the properties of the $\pi(1300)$ and $\pi(1800)$ are likely to be sensitive to details of the long-range part of the quark-quark interaction because the constituent-quark wave functions will possess material support at large interquark separation.  Hence the development of an understanding of their properties may provide information about light-quark confinement, which complements that obtained via angular momentum excitations\,\cite{a1b1}.  

That Eq.\,(\ref{gengmor}) is a powerful result is further emphasised by the fact that it is applicable here, too \cite{krassnigg1,krassnigg2}.  The result holds at each pole common to the pseudoscalar and axial-vector vertices and therefore it also impacts upon the properties of non-ground-state pseudoscalar mesons.  Let's work with a label $n\geq 0$ for the pseudoscalar mesons: $\pi_n$, with $n=0$ denoting the ground state, $n=1$ the state with the next lowest mass, and so on.  By assumption, $m_{\pi_{n\neq 0}}>m_{\pi_0}$, and hence $m_{\pi_{n\neq 0}} > 0$ in the chiral limit.  In addition
\begin{equation}
0 < \rho_{\pi_{n}}^0(\zeta):= \lim_{\hat m\to 0} \rho_{\pi_{n}}(\zeta) < \infty \,, \; \forall \, n\,.
\end{equation}
Hence, it is a necessary consequence of chiral symmetry and its dynamical breaking in QCD; viz., Eq.\,(\ref{gengmor}), that
\begin{equation}
\label{fpinzero}
f_{\pi_n}^0 \equiv 0\,, \forall \, n\geq 1\,.
\end{equation}
This result means that in the presence of DCSB all pseudoscalar mesons except the ground state decouple from the weak interaction.  NB.\ Away from the chiral limit the quantities $f_{\pi_n}$ alternate in sign; i.e., they are positive for even $n$ but negative for odd $n$.  This is an essential prediction of spectral positivity in quantum field theory and follows because $f_{\pi_n}$ are the residues of colour-singlet poles in a \textit{vertex} that, considered as a function of $P^2$, is continuous and does not vanish between adjacent bound states.

These arguments are legitimate in any theory with a valid chiral limit.  It is logically possible that such a theory does not exhibit DCSB; i.e., realises chiral symmetry in the Wigner-Weyl mode.  Equation (\ref{gengmor}) is still valid in the Wigner phase.  However, its implications are different; namely, in the Wigner phase, one has 
\begin{equation}
\label{Bm0}
B^W(0,\zeta^2)\propto m(\zeta) \propto\hat m\,;
\end{equation}
i.e., the mass function and constituent-quark mass vanish in the chiral limit.  Equations (\ref{bwti}) apply if there is a massless bound state in the chiral limit.  Suppose such a bound state persists in the absence of DCSB.\footnote{If that is false then considering this particular case is unnecessary.  However, it is true at the transition temperature in QCD \cite{bastirev}.} It then follows from Eqs.\,(\ref{bwti}) \& (\ref{Bm0}) that 
\begin{equation}
f^W_{\pi_0} \propto \hat m\,.
\end{equation}
In this case the leptonic decay constant of the ground state pseudoscalar also vanishes in the chiral limit, and hence all pseudoscalar mesons are blind to the weak interaction.

As further examples, exact results have also been established for: $\pi \pi$ scattering \cite{bicudo,marisbicudo}; the $\gamma \pi^0_{n=0} \gamma$ \cite{kekez} and $\gamma \pi^0_{n\geq 1} \gamma$ \cite{krassnigg2} transition form factors; and the $\gamma \pi\pi\pi$ transition form factor \cite{mariscotanch2}.

\medskip

%
\hspace*{-\parindent}\underline{\textit{Predictive tool}}\\[1ex]
It is now recognised that the leading-order term in the systematic, nonperturbative symmetry-preserving truncation of the DSEs is provided by the renormalisation-group-improved rainbow-ladder truncation, which has been used widely; e.g., Refs.\,\cite{mr97,jainmunczek,klabucar} and references thereto.  A practical renormalisation-group-improved rainbow-ladder truncation preserves the one-loop ultraviolet behaviour of perturbative QCD.  However, a model assumption is required for the behaviour of the kernel in the infrared; viz., on the domain $Q^2 \lsim 1\,$GeV$^2$, which corresponds to length-scales $\gsim 0.2\,$fm.  This is the confinement domain whereupon little is truly known about the interaction between light-quarks.  That information is, after all, what we seek.  The application of a single model to an extensive range of JLab-related phenomena is reviewed in Ref.\,\cite{pieterrev} and summarised in Sec.\,5.2.2 of Ref.\,\cite{HUGS}.  Herein we simply note that the one-parameter renormalisation-group-improved rainbow-ladder model introduced in Ref.\,\cite{maristandy1} provides an excellent tool with which to illustrate exact results, such as those described above, and moreover has proved to be a valuable predictive device \cite{maristandy3,volmer}.

\section{Baryon Properties}
While the significant progress made with the study of mesons is good, it does not directly impact on the important challenge of baryons.  Mesons fall within the class of two-body problems.  They are the simplest bound states for theory.  However, the absence of meson targets poses significant difficulties for the experimental verification of predictions such as those reported above.  On the other hand, it is relatively straightforward to construct a proton target but, as a three-body problem in relativistic quantum field theory, here the difficulty is for theory.  With this problem the current expertise is approximately at the level it was for mesons ten years ago; namely, model building and phenomenology, making as much use as possible of the results and constraints outlined above.  

Modern, high-luminosity experimental facilities that employ large momentum transfer reactions are providing remarkable and intriguing new information on nucleon structure \cite{gao,leeburkert}.  For an example one need only look so far as the discrepancy between the ratio of electromagnetic proton form factors extracted via Rosenbluth separation and that inferred from polarisation transfer \cite{jones,roygayou,gayou,arrington,qattan}.  This discrepancy is marked for $Q^2\gsim 2\,$GeV$^2$ and grows with increasing $Q^2$.  At such values of momentum transfer, $Q^2 > M^2$, where $M$ is the nucleon's mass, a veracious understanding of these and other contemporary data require a Poincar\'e covariant description of the nucleon.  

A natural primary aim is to develop a good theoretical picture of the proton's electromagnetic form factors.  To this end Ref.\,\cite{hoell} proposed that the nucleon is at heart composed of a dressed-quark and nonpointlike diquark.  One element of that study is the dressed-quark propagator.  The form used expresses the features described above and carries no free parameters, because its behaviour was fixed in analyses of meson observables \cite{mark}.  The nucleon bound state was subsequently realised via a Poincar\'e covariant Faddeev equation, which incorporates scalar and axial-vector diquark correlations.  In this there are two parameters: the mass-scales associated with the correlations.  They were fixed by fitting to specified nucleon and $\Delta$ masses: the values are listed in Table~\ref{ParaFix}.  The study thus arrived at a representation of the nucleon that possesses no free parameters with which to influence the nucleons' form factors.

\begin{center}
\begin{table}[t]
\begin{tabular*}{1.0\textwidth}{
c@{\extracolsep{0ptplus1fil}}c@{\extracolsep{0ptplus1fil}}
|c@{\extracolsep{0ptplus1fil}}c@{\extracolsep{0ptplus1fil}} |c@{\extracolsep{0ptplus1fil}}c@{\extracolsep{0ptplus1fil}}}
\hline
$M_N$ & $M_{\Delta}$~ & $m_{0^{+}}$ & $m_{1^{+}}$~ &
$\omega_{0^{+}} $ & $\omega_{1^{+}}$ \\
1.18 & 1.33~ & 0.79 & 0.89~ & 0.56=1/(0.35\,{\rm fm}) & 0.63=1/(0.31\,{\rm fm}) \\
\hline
\end{tabular*}
\caption{Mass-scale parameters (in GeV) for the scalar and axial-vector diquarks, fixed by fitting nucleon and $\Delta$ masses: the fitted mass was offset to allow for ``pion cloud'' contributions \protect\cite{hechtfe}, which reduce both the nucleon and $\Delta$ masses to their experimental values.  $\omega_{J^{P}}= m_{J^{P}}/\surd 2$ is the width-parameter in the nonpointlike $(qq)_{J^P}$-diquark's Bethe-Salpeter amplitude: its inverse is a gauge of the diquark's matter radius.  Charge radii are estimated in Ref.\,\cite{marisdqrad}. (Adapted from Ref.\,\protect\cite{hoell}.)\label{ParaFix}}
\end{table}
\end{center}

At this point only a specification of the nucleons' electromagnetic interaction remained.  Its formulation was primarily guided by a requirement that the nucleon-photon vertex satisfy a Ward-Takahashi identity.  The interaction depends on three parameters tied to properties of the axial-vector diquark correlation: $\mu_{1^+}$ and $\chi_{1^+}$, respectively, the axial-vector diquarks' magnetic dipole and electric quadrupole moments; and $\kappa_{\cal T}$, the strength of electromagnetic axial-vector $\leftrightarrow$ scalar diquark transitions.  Calculated results for the nucleons' form factors, however, were not materially sensitive to these parameters \cite{hoell}, which enabled a prediction to be made \cite{hoell2}: $\mu_p\, G_E^p(Q^2)/G_M^p(Q^2)=0$ at $Q^2 \simeq 6.5\,$GeV$^2$; namely, at the point for which $G_E^p(Q^2)=0$.  The behaviour of $\mu_p\, G_E^p(Q^2)/G_M^p(Q^2)$ owes itself primarily to spin-isospin correlations in the nucleon's Faddeev amplitude.  An experiment is planned at JLab that will acquire data on this ratio to $Q^2=9.0\,$GeV$^2$ \cite{E04108}.  It is expected to begin running around the beginning of 2008.

This framework can naturally be applied to calculate weak and strong form factors of the nucleon.  Preliminary studies of this type are reported in Refs.\,\cite{blochgA,oettelgA}.  Such form factors are sensitive to different aspects of quark-nuclear physics and should prove useful, e.g., in constraining coupled-channel models for medium-energy production reactions on the nucleon.  

We will briefly describe first results for three such form factors: the axial-vector and pseudoscalar nucleon form factors, which appear in the axial-vector--nucleon current
\begin{equation}
\label{5muNN}
J_{5\mu}^j(P^\prime,P) = i \bar u(P^\prime) \frac{\tau^j}{2} \Lambda_{5\mu}(q;P) u(P)
= \bar u(P^\prime) \gamma_5 \frac{\tau^j}{2} \left[ \gamma_\mu \, g_A(q^2) + q_\mu\, g_P(q^2) \right] u(P)\,,
\end{equation}
where $q=P^\prime - P$, $j=1,2,3$ is the isospin index, and the nucleon spinor, $u(P)$, is defined in Ref.\,\cite{hoell}; and the pion-nucleon coupling
\begin{equation}
\label{Jpi}
J^j_\pi(P^\prime,P) = \bar u(P^\prime) \Lambda_{\pi}^j(q;P) u(P)
= g_{\pi NN}(q^2) \bar u(P^\prime) i \gamma_5 \tau^j u(P)\,.
\end{equation} 

\begin{figure}[t]
\includegraphics*[width=0.58\textwidth,angle=-90]{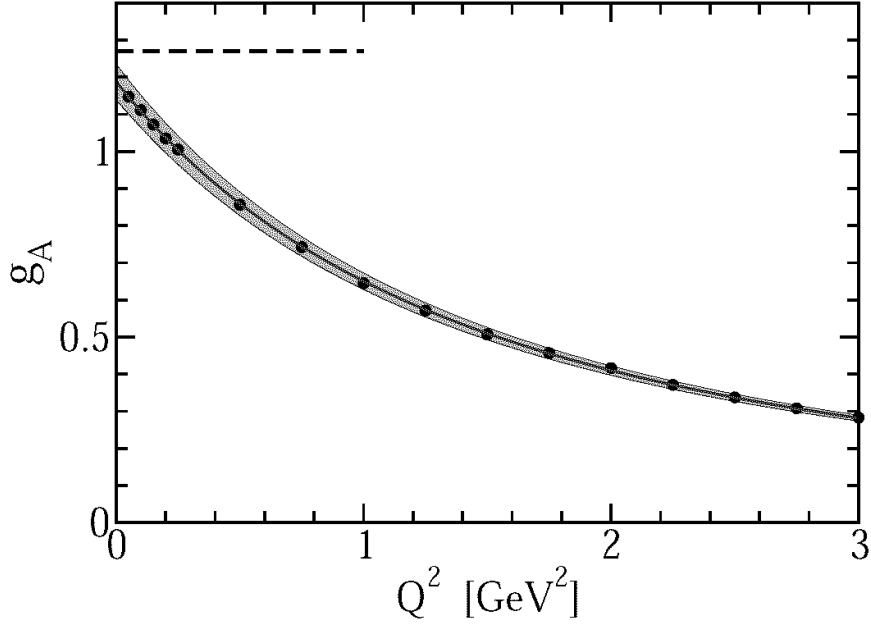}
\caption{\label{F1} \textit{Filled circles}: $g_A(Q^2)$ in Eq.\,(\protect\ref{5muNN}) calculated in the chiral limit using the nucleon Faddeev amplitudes and the axial-vector-nucleon vertex obtained from Eqs.\,(\protect\ref{avansatz}), (\protect\ref{vA1}) \& (\protect\ref{vA01}).  \textit{Solid line}: dipole fit to the calculation, with mass-scale $m_D^A=1.69\,$GeV.  The shaded band delimits the result's variation subject to $10\,$\% changes in the parameter values in Eq.\,(\protect\ref{paramsK}).  The experimental value of the nucleon's axial coupling ($g_A \approx 1.27$) is marked by a dashed line.}
\end{figure}

\begin{figure}[t]
\begin{minipage}[c]{0.65\textwidth}
\hspace*{-5ex}
\includegraphics*[width=0.95\textwidth,angle=-90]{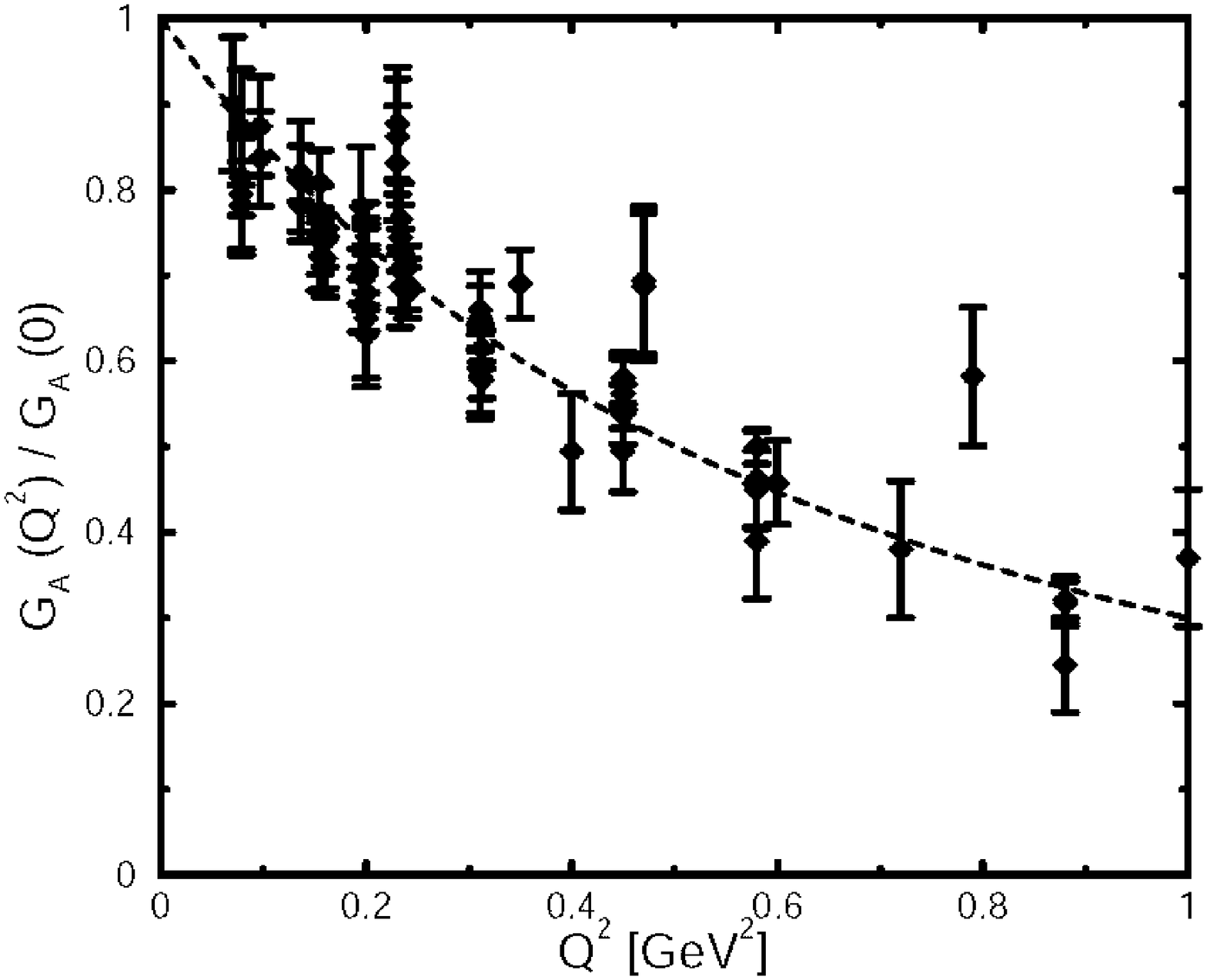}\\[-48.4ex]
\hspace*{-2.0ex}
\includegraphics*[width=1.01\textwidth,angle=-90]{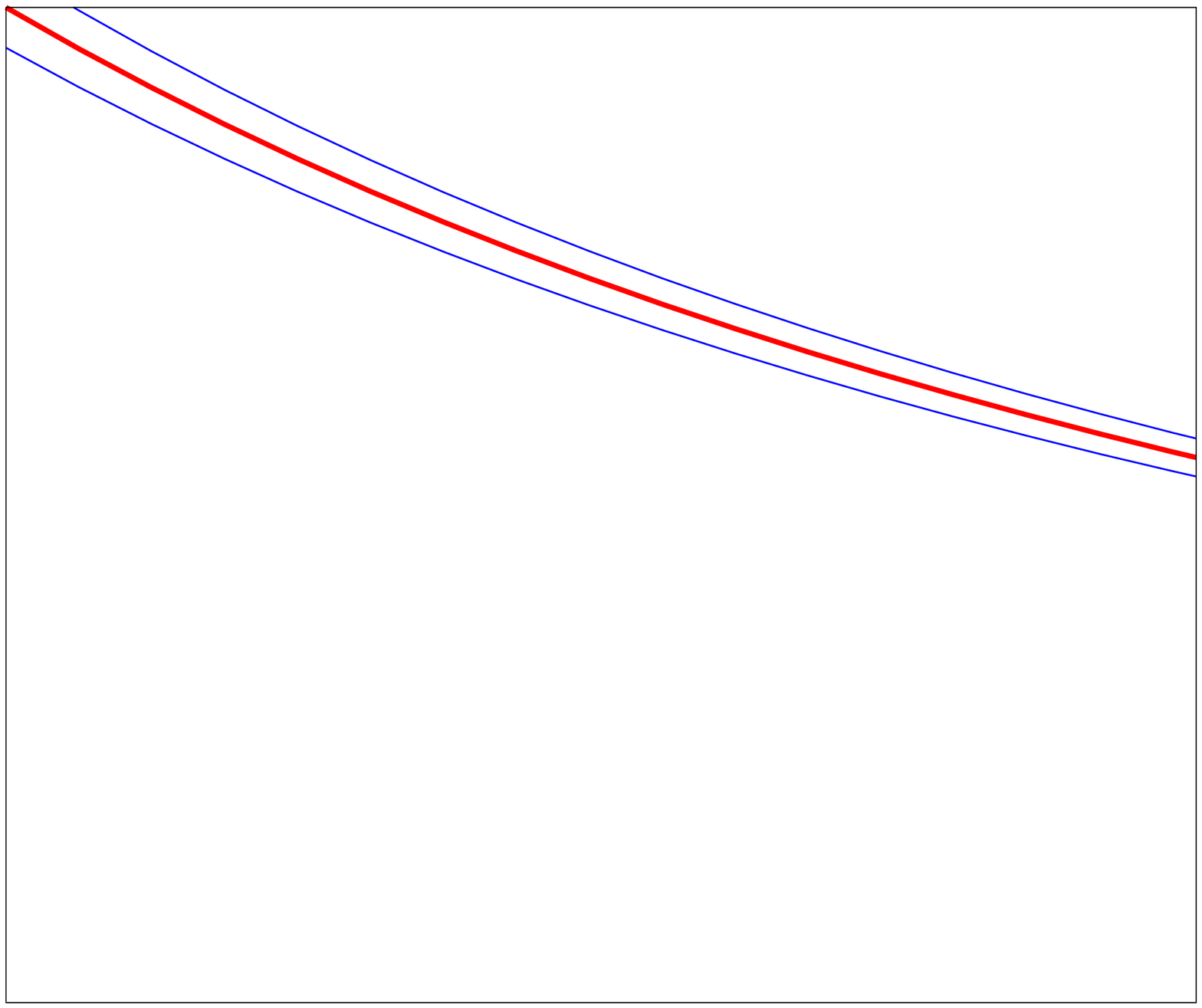}
\end{minipage}
\caption{\label{F2} Calculated chiral-limit result for $g_A(Q^2)/g_A(0)$, \textit{solid line}, compared with data obtained via pion electroproduction in the threshold region, as described in Ref.\,\protect\cite{bernard}.  \textit{Dashed line}: dipole fit to data with mass-scale $m_D^{AE}=1.1\,$GeV.}
\end{figure}

\begin{figure}[t]
\includegraphics*[width=0.58\textwidth,angle=-90]{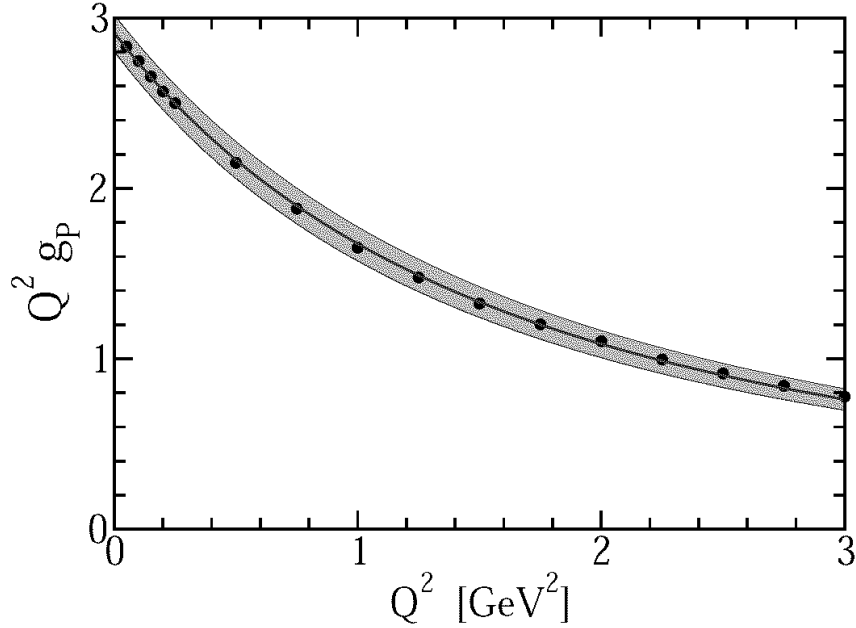}
\caption{\label{F3} \textit{Filled circles}: Chiral limit result for $Q^2 g_P(Q^2)$ in Eq.\,(\protect\ref{5muNN}) calculated as described in the caption of Fig.\,\protect\ref{F1}.  \textit{Solid line}: dipole fit to the calculation, with mass-scale $m_D^P=1.77\,$GeV.  The shaded band delimits the result's variation subject to $10\,$\% changes in the parameter values in Eq.\,(\protect\ref{paramsK}).}
\end{figure}

\begin{figure}[t]
\includegraphics*[width=0.6\textwidth,angle=-90]{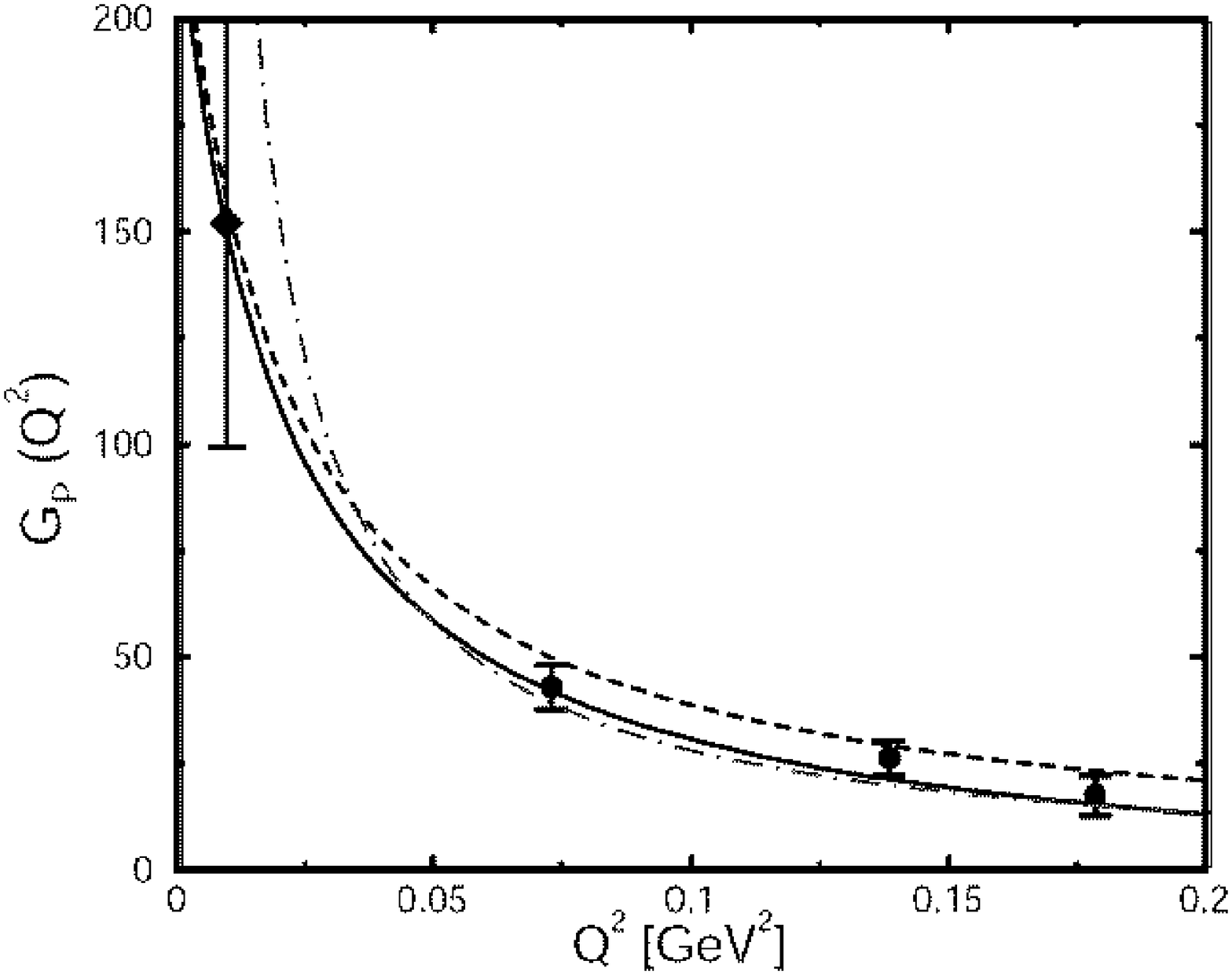}
%
\caption{\label{F4} Chiral-limit result for $g_P(Q^2)$, \textit{dash-dot curve}.  Data obtained via pion electroproduction (\textit{filled circles}) \protect\cite{choi} and world average for muon capture at $Q^2=0.88 m_\mu^2$ (\textit{filled diamond}).  \textit{Dashed curve} -- current-algebra result; and  \textit{solid curve} -- next-to-leading order chiral
perturbation theory result \protect\cite{bernard}.}
\end{figure}

In the chiral limit the pseudovector vertex of Eq.\,(\ref{5muNN}) takes the following form in the neighbourhood of $q^2=0$ \cite{mrt98}
\begin{equation}
\Lambda^j_{5\mu}(q;P) \stackrel{q^2\sim 0}{=} \mbox{regular}\; + \frac{q_\mu}{q^2} f_\pi \Lambda_\pi^j(q;P)\,,
\end{equation}
where $\Lambda_\pi^j(q;P)$ is the pion-nucleon vertex and ``regular'' denotes non-pole terms.  In addition, $q_\mu J_{5\mu}^j(P^\prime,P) =0$.  From these observations ensues the Goldberger-Treiman relation:
\begin{equation}
M \, g_A(q^2=0) = f_\pi \,g_{\pi NN}(q^2=0)\,,
\end{equation}
where $M$ is the calculated nucleon mass and $g_A(q^2)$ is solely associated with the regular part of the axial-vector vertex.

The calculation of electromagnetic form factors sets a pattern for determining $g_A(q^2)$, $g_P(q^2)$ and $g_{\pi NN}(q^2)$, and that is what we follow.  We need to know how a dressed-quark couples to an axial-vector probe.  In the chiral limit the dressed-quark--axial-vector vertex satisfies Eq.\,(\ref{avwtim}) with the mass-dependent term omitted.  Hereafter we'll assume isospin symmetry so that $S_u=S_d$, in which case the chiral-limit axial-vector Ward-Takahashi identity is solved by 
\begin{equation}
\label{avansatz}
\Gamma_{5\mu}^j(k;Q) = \gamma_5 \frac{\tau^j}{2} \left[ \gamma_\mu \Sigma_A(k^2_+,k_-^2) + 2 k_\mu \gamma\cdot k \Delta_A(k^2_+,k_-^2) + 2\,i\, \frac{Q_\mu}{Q^2}\Sigma_B(k^2_+,k_-^2)   \right],
\end{equation}
with 
\begin{equation}
\Sigma_F(\ell_1^2,\ell_2^2) = \sfrac{1}{2}\,[F(\ell_1^2)+F(\ell_2^2)]\,,\;
\Delta_F(\ell_1^2,\ell_2^2) =
\frac{F(\ell_1^2)-F(\ell_2^2)}{\ell_1^2-\ell_2^2}\,,
\label{DeltaF}
\end{equation}
where $F= A, B$; viz., the scalar functions in Eq.\,(\ref{Sgeneral}).  Naturally, Eq.\,(\ref{avansatz}) is not a unique \textit{Ansatz} for the dressed-quark--axial-vector vertex but it is an adequate starting point.  

For the pion-nucleon coupling, one needs the pion's Bethe-Salpeter amplitude and its extension off pion mass-shell.  In chiral QCD we have Eqs.\,(\ref{bwti}), upon which we base the \textit{Ansatz}
\begin{equation}
\label{piBSA}
\Gamma_\pi^j(k;Q) =  i \gamma_5 \tau^j \, \sfrac{1}{{\cal N}_\pi}\, \Sigma_B(k_+^2,k_-^2)\,,
\end{equation}
where ${\cal N}_\pi$ is the canonical normalisation constant calculated with this amplitude.  (See, for example, Eqs.\,(37) \& (38) of Ref.\,\protect\cite{hoell}.)

We also need to know the following vertices: pion--axial-vector-diquark;  axial-vector-probe--axial-vector-diquark; and the pion- and axial-vector-probe-induced scalar-diquark $\leftrightarrow$ axial-vector-diquark transitions.  For these we follow Ref.\,\cite{oettelgA}:
\begin{eqnarray}
\Gamma^{\pi 1}_{\alpha\beta}(p^\prime,p) & = & \frac{\kappa^{\pi 1}}{2 M_N} \, \frac{M_Q^E}{f_\pi}\, \epsilon_{\alpha\beta\mu\nu} (p^\prime+p)_\mu Q_\nu\,, 
\label{pi1}\\
\Gamma^{A 1}_{\mu\alpha\beta}(p^\prime,p) & = & \sfrac{1}{2} \kappa^{A 1} \, \epsilon_{\mu\alpha\beta\nu} (p^\prime+p)_\nu + 2 f_\pi \frac{Q_\mu}{Q^2} \, \Gamma^{\pi 1}_{\alpha\beta}(p^\prime,p)\,,\label{vA1}\\
\Gamma^{\pi 01}_\beta(p^\prime,p) & = & -i \kappa^{\pi 01} \, \frac{M_Q^E}{f_\pi}\, Q_\beta\,,\label{pi01}\\
\Gamma_{\mu\beta}^{A 01}(p^\prime,p) & = & i M_N \kappa^{A 01} \delta_{\mu\beta} + 2 f_\pi \frac{Q_\mu}{Q^2} \, \Gamma^{\pi 01}_{\beta}(p^\prime,p)\,,
\label{vA01}
\end{eqnarray}
where $M_Q^E$ is the Euclidean light-quark constituent-mass \cite{mr97}, $p$ \& $p^\prime$ are the incoming and outgoing diquark momenta and $Q=(p^\prime-p)$.  Each  \textit{Ansatz} introduces one parameter, for which typical values are \cite{oettelgA}:
\begin{equation}
\label{paramsK}
\kappa^{\pi 1} \simeq  \kappa^{A 1} \simeq 4.5\,,\; \kappa^{\pi 01} \simeq 3.9\,, \; \kappa^{A 01} \simeq 2.1\,.
\end{equation}
We used these to obtain the results reported below, with the bands representing a variation of $\pm 10\,$\%.  NB.\ A scalar diquark does not couple to a single pseudoscalar or axial-vector probe.

With the elements heretofore described we have an analogue of the top four diagrams in Fig.\,1 of Ref.\,\cite{hoell}.  This is necessary but not sufficient to guarantee that the axial-vector--nucleon vertex automatically fulfills the chiral Ward-Takahashi identity for on-shell nucleons.  Work on improvement is underway.

In Fig.\,\ref{F1} we display our result for the nucleon's axial-vector form factor.  A comparison with extant data is provided in Fig.\,\ref{F2}.  We attribute the mismatch to a failure of the axial-vector-nucleon vertex obtained from Eqs.\,(\protect\ref{avansatz}), (\protect\ref{vA1}) \& (\protect\ref{vA01}) to properly express the diquarks' nonpointlike nature: the result is thus too hard.

In Fig.\,\ref{F3} we depict our result for the nucleon's induced-pseudoscalar form factor.  A comparison with data is provided in Fig.\,\ref{F4}.  The form factor is dominated by the pion pole in the neighbourhood of $q^2= -m_\pi^2$, which for our chiral-limit calculation is $q^2\sim 0$.  In this case the comparison with data is more favourable, particularly once one allows for a shift of the pion pole to $q^2=0$ in our chiral-limit calculation.  We attribute this to Eqs.\,(\ref{piBSA}), (\ref{pi1}) \& (\ref{pi01}); viz., as it is based on Eqs.\,(\ref{bwti}), our calculation incorporates a fairly accurate representation of pion structure and the pion nucleon coupling.

\begin{figure}[t]
\includegraphics*[width=0.58\textwidth,angle=-90]{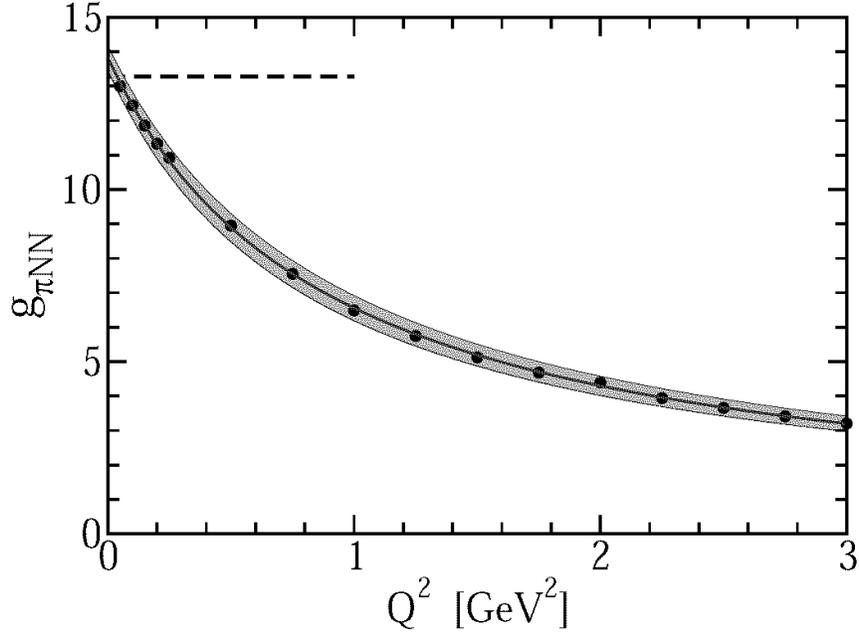}
\caption{\label{F5} \textit{Filled circles}: Chiral limit result for $g_{\pi NN}(Q^2)$ in Eq.\,(\protect\ref{Jpi}) calculated using the nucleon's Faddeev amplitude and the $\pi NN$ vertex constructed from Eqs.\,(\ref{piBSA}), (\ref{pi1}) \& (\ref{pi01}).  \textit{Solid line}: monopole fit to the calculation, with mass-scale $m_M^\pi=0.95\,$GeV.  The shaded band delimits the result's variation subject to $10\,$\% changes in the parameter values in Eq.\,(\protect\ref{paramsK}).  The experimental value of the $\pi NN$ coupling ($g_{\pi NN} \approx 13.4$) is marked by a dashed line.}
\end{figure}

This view is supported by our result for $g_{\pi NN}(q^2)$, which is depicted in  Fig.\,\ref{F5}.  Within reasonable variation of the parameters that characterise the pion-nucleon vertex, the calculated value of $g_{\pi NN}^0(0)$ is consistent with standard phenomenology.  Our result yields a chiral-limit value $r_{\pi NN}^0 \simeq 0.51\pm 0.02 \,$fm.  For comparison, a massive-quark value of $r_{\pi NN}\sim 0.3\,$fm appears in Ref.\,\protect\cite{machleidt}, while $r_{\pi NN}\sim 0.93\,$--$\,1.06\,$fm is employed in Ref.\,\cite{lee}. 

In order to improve upon these preliminary results, construction must be completed of an axial-vector--nucleon vertex that automatically fulfills the chiral Ward-Takahashi identity for on-shell nucleons described by the solution of the Faddeev equation.  This will subsequently lead to an improved pion-nucleon vertex.  In addition, as is known to be necessary for an accurate description of nucleon electromagnetic properties, the effect of pseudoscalar meson loops on the axial and pseudoscalar couplings must be incorporated.  These steps are prerequisites for the reliable extension of our Poincar\'e covariant model to weak and pionic processes.

\section{Epilogue}
The perturbative formulation of QCD fails spectacularly to account for even the simplest bulk properties of hadrons.  Two fundamental, emergent phenomena are responsible: confinement and dynamical chiral symmetry breaking.  Their importance is difficult to overestimate.  

Dynamical chiral symmetry breaking (DCSB) is a singularly effective mass generating mechanism.  It takes the almost massless light-quarks of perturbative QCD and converts them into the massive constituent-quarks whose mass sets the scale which characterises the spectrum of the strong interaction.  The phenomenon is understood via QCD's gap equation, whose solution delivers a mass function with a momentum-dependence that connects the perturbative and nonperturbative-constituent-quark domains.  

Despite the fact that light-quarks are made heavy, the mass of the pseudoscalar mesons remains peculiarly small.  That, too, owes to DCSB, expressed this time in a remarkable relationship between QCD's gap equation and those colour singlet Bethe-Salpeter equations which have a pseudoscalar projection.  Goldstone's theorem is a natural consequence of this connection.

The Dyson-Schwinger equations (DSEs) provide a natural framework for the exploration of QCD's emergent phenomena.  They are a generating tool for perturbation theory and thus give a clean connection with processes that are well understood.  Moreover, they admit a systematic, symmetry preserving and nonperturbative truncation scheme, and thereby give access to strong QCD in the continuum.  On top of this, a quantitative feedback between DSE and lattice-QCD studies is today proving fruitful.

The existence of a sensible truncation scheme enables the proof of exact results using the DSEs.  That the truncation scheme is also tractable provides a means by which the results may be illustrated, and furthermore a practical tool for the prediction of observables that are accessible at contemporary experimental facilities.  The consequent opportunities for rapid feedback between experiment and theory brings within reach an intuitive understanding of nonperturbative strong interaction phenomena.

An important challenge is the study of baryons.  Modern, high-luminosity experimental facilities employ large momentum transfer reactions to probe baryon structure, and they are providing remarkable and intriguing new information.  A true understanding of much contemporary data requires a Poincar\'e covariant description of the nucleon.  This can be obtained with a Faddeev equation that describes a baryon as composed primarily of a quark core, constituted of confined quark and confined diquark correlations, but augmented by pseudoscalar meson cloud contributions that are sensed by long wavelength probes.  Short wavelength probes pierce the cloud, and expose spin-isospin correlations and quark orbital angular momentum within the baryon.  The veracity of the elements in this description makes plain that a picture of baryons as a bag of three constituent-quarks is profoundly misleading.


\begin{theacknowledgments}
CDR expresses his deep gratitude for the hospitality and support of the organisers of the \textit{X$^{\rm th}$ Mexican Workshop on Particles and Fields}.
In preparing this article we benefited from conversations with A.~Bashir, P.~Jaikumar, A.~Krassnigg, P.~Maris, A.~Raya, and P.\,C.~Tandy.
This work was supported by: 
Department of Energy, Office of Nuclear Physics, contract no.\ W-31-109-ENG-38; 
\textit{Helmholtz-Gemeinschaft} Virtual Theory Institute VH-VI-041; 
the \textit{A.\,v.\ Humboldt-Stiftung} via a \textit{F.\,W.\ Bessel Forschungspreis}; 
and benefited from the facilities of ANL's Computing Resource Center.
\end{theacknowledgments}


\end{document}